\documentclass[preprint,12pt]{elsarticle}

\usepackage{amssymb}
\usepackage{amsmath}
\usepackage{makeidx}         % allows index generation
\usepackage{graphicx}        % standard LaTeX graphics tool
\usepackage{color}

\newcommand{\beq}{\begin{eqnarray}}
\newcommand{\eeq}{ \end{eqnarray} }
\newcommand{\bit}{\begin{itemize}}
\newcommand{\eit}{\end{itemize}}
\newcommand{\p}{\partial}

\newcommand{\bt}{\mathbf{t}}
\newcommand{\bn}{\mathbf{n}}

\newcommand{\bp}{{\bf p}}

\journal{Physica D}

\begin{document}

\begin{frontmatter}



\title{Membrane tension  feedback on shape and motility of eukaryotic  cells}


\author[freib]{Benjamin Winkler}
\author[arg,northw]{Igor S. Aranson\fnref{tele1}}\ead{aronson@anl.gov}
\author[freib]{Falko Ziebert\fnref{tele2}}\ead{falko.ziebert@physik.uni-freiburg.de}
\fntext[tele1]{phone: +1 630 252 9725; fax: +1 630 252 7777}
\fntext[tele2]{phone: +49 761 203 97779; fax: +49 761 203 5855}

\address[freib]{Physikalisches Institut, Albert-Ludwigs-Universit\"at, 79104 Freiburg, Germany}
\address[arg]{Materials Science Division, Argonne National Laboratory, 9700 S. Cass Avenue,
Argonne, IL 60439, USA}
\address[northw]{Engineering Sciences and Applied Mathematics, Northwestern University, 2145
Sheridan Road, Evanston, IL 60202, USA}

\begin{abstract}
In the framework of a phase field model of a single cell crawling on a substrate, 
we investigate how  the properties of the cell membrane affect the shape and motility of the cell. 
Since the membrane influences the cell dynamics on multiple levels and provides a nontrivial feedback, 
%to its motility and shape,  
we consider the following fundamental interactions: 
(i) the reduction of  the actin polymerization rate by membrane tension; 
(ii) area conservation of the cell's two-dimensional cross-section
vs.~conservation of its circumference (i.e.~membrane inextensibility); 
and (iii) the contribution from  the membrane's bending energy to the shape and integrity of the cell.
As in experiments, we investigate two pertinent observables -- the cell's velocity and its aspect ratio. 
We find that the most important effect is the feedback of membrane tension on the actin polymerization.
Bending rigidity has only minor effects, visible mostly in dynamic reshaping events, as exemplified
by collisions of the cell with an obstacle.
%{\it \color{red}In addition,  we study the interaction of a cell with an obstacle or a hole/tunnel,
%where the membrane tension feedback and bending stiffness increase the 
%stability/cohesion of the cell. }
\end{abstract}


\end{frontmatter}


\section{Introduction}
\label{intro}

Motility of cells crawling on substrates attracts substantial 
interest among biologists, physicists, and material scientists alike.
Cell motility  is a fundamental phenomenon that is crucial for a variety 
of biological processes, from  morphogenesis to immune response.  
It is also  involved in pathologies like cancer growth and metastasis. 
Like swimming microorganisms, crawling  motile cells  
are  natural and interesting realizations of active, self-propelled systems, 
displaying self-organized dynamics, flows, as well as intriguing  collective effects. 
Moreover, motile  cells and living tissues are  inspiring novel adaptive  materials 
with intricate properties like active visco-elastic response and self-healing. Cellular materials,  
responding  to  the topography, elasticity, and surface chemistry of the substrate they are in contact with, 
currently inspire microstructured design strategies for cell sorting and guiding.

The main processes involved in the motion of eukaryotic cells (such as 
keratocytes, fibroblasts or neutrophils) are the following: the generation of
a propulsive force by actin polymerization against the cell's membrane, 
 %intermittent or maturing 
the formation of adhesive contact to the substrate to transfer this propulsion force and
to move forward, and finally,  the action of molecular motors in 
determining the cell's polarity and to retract the rear of the cell \cite{Abercrombie}. 
All these processes have been modeled in some detail, 
and  models for whole moving cells have been recently developed
\cite{Kruse:2006.1,WolgemuthBP,shao2010,ziebert2011model,recho2013}.
However, there is another important player in the game, that has been neglected
(or its consequences not yet thoroughly studied) in most of the modeling approaches: 
namely, the membrane enclosing the cell.
The cell membrane  represents a movable interface which constitutes an intricate theoretical and numerical 
problem. In addition, 
membrane tension leads to a global force feedback, affecting  the 
propulsion by ratcheting the actin filaments. Moreover, membrane bending rigidity may be relevant
in some cases, especially for cell collisions with other cells or obstacles. 

The first detailed experimental study on the effects of membrane tension on spreading 
cells (fibroblasts) dates back no longer than in 2000 \cite{Raucher_Sheetz}.
There, an inverse relation between spreading/lamellipodium extension and
membrane tension was found: lowering the membrane tension by adding detergents (deoxycholic acid)
or lipids led to an increased spreading and extension, while an increase in tension
by placing cells in a hypotonic medium reduced both effects.
The authors concluded that membrane tension may constitute a global coupling 
%(but still modulated via the curvature)
involved in determining both the cell's shape and the propulsion dynamics, 
cf.~also the recent reviews \cite{Sheetzrev,Fletcherrev}.
The effect of membrane tension was studied also for neutrophils, both 
during pseudopod formation and for fully developed motion \cite{Houk}, 
for spreading fibroblasts \cite{Gauthier_Sheetz}, as well as 
for moving keratocytes \cite{Keren_mb}.  Some of the observed effects include: 
(i) increased membrane tension can cause leukocytes to stop moving \cite{Houk}; 
(ii) reducing tension can stimulate  moving keratocytes to develop several fronts \cite{Keren_mb}; 
(iii) softening the cell membrane does not affect the velocity of keratocytes \cite{Keren_mb,Fuhs_Kaes},
it only increases the retrograde flow of actin towards the cell's interior.

Membrane tension  has been recently taken  into account, for instance,  in 
the one-dimensional model for growth cones \cite{CraigMogil}, 
as well as the steady state.  The force balance-based model in \cite{Schweitzer}
includes also explicit adhesion dynamics between the actin cortex and the membrane. 
Very recently, tension gradients and flows inside
the membrane were  addressed \cite{LieberKeren,MogilPONE}.
%as well as \cite{LieberKeren}. Very recent, maybe not relevant: \cite{Itoh}.
However, these models do not take shape changes
% -- and hence the overall feedback of the membrane -- 
into account, obviously an important aspect of the membrane's feedback. As a result,  they
can not properly describe the onset/cessation of motion. These two important aspects can be 
easily and inherently modeled within the phase field approach recently developed for 
motile cells \cite{shao2010,ziebert2011model,shaoPNAS,ziebert2013effects,
lober2014modeling,ziebertEPJST,CatesNCom,LoeberSciRep}, self-propelled  active droplets \cite{CatesPNAS,VoigtJRSI,yoshinaga2014spontaneous} and  synthetic polymeric capsules \cite{kolmakov2012designing}. 
Here we include and study the most pertinent membrane effects -- tension and its feedback 
on polymerization, as well as bending. The study is performed  within a simple phase 
field approach for a moving cell. 

\section{Phase field model for a crawling cell}
\label{model}

The phase field approach to cell motility has been recently reviewed  in \cite{ebook}.
Instead of modeling the cell's interface (i.e.~the membrane) explicitly, an auxiliary field,
the phase field $\rho(x,y;t)$, is introduced. It evaluates to $\rho=1$ within the cell
and to $\rho=0$ outside the cell, with a smooth transition region in between describing the
`smeared' interface.
The simplest implementation of the phase field approach is via a scalar order parameter equation 
\beq\label{Fdef}
\p_t\rho=-\frac{\delta F_P}{\delta\rho}\,\,,\,\,\,{\rm where}\,\,\,
%\eeq
%where
%\beq
F_p=\int\left[f(\rho)+D_\rho(\nabla\rho)^2\right]dx\,dy\,.
\eeq
Here $f(\rho)=\frac{\rho^2(1-\rho)^2}{4}$ is a double well potential with minima at $\rho=0$ and $\rho=1$
(the two `phases').
The phase field free energy $F_p$ in addition includes a surface energy term penalizing 
%the length of 
interfaces. Equation (\ref{Fdef}) yields 
\beq\label{simple}
\p_t\rho=D_\rho\Delta\rho-\rho(1-\rho)(\delta-\rho)=:\lambda\,,
\eeq
where $\delta=\frac{1}{2}$ is the `pressure difference' between the two `phases'. % ($\rho=1$ and $\rho=0$). 
For $\delta=\frac{1}{2}$ the free energy of both phases is equal, and hence a planar interface  
connecting states $\rho=0$ and $\rho=1$ is stationary. In case $\delta$ deviates from
this value, the interface moves either forward or backward, i.e.~the cell expands or retracts.

%Note that the 2nd term is $(-)f'(\rho)$.

We used this simple framework to model a moving cell \cite{ziebert2011model} 
by coupling the phase field Eq.~(\ref{simple})  to the polarization field $\bp$, describing
the averaged local orientation of the actin filaments inside the cell:
\begin{eqnarray} \label{reqsimple0}
\partial _t\rho&=&D_\rho\Delta\rho-\rho(1-\rho)(\delta-\sigma|\bp|^2-\rho)-\alpha\bp \cdot\nabla\rho\,,\\
\label{peqsimple}
\partial_t\bp&=&D_p\Delta\bp
-\beta \nabla\rho
%-\beta g\left(\nabla\rho\right)
\,-\tau_1^{-1}\bp   -\tau_2^{-1}(1-\rho^2)\bp 
-\gamma\left[(\nabla\rho)\cdot\bp\right]\bp  \,.
\end{eqnarray}
%$g\left(z\right) =\frac{z}{\sqrt{1+\epsilon z^{2}}}$
%limits polymerization $\propto\beta$ for steep gradients.
In this description, the $\alpha$-term models the propulsion of the cell's interface by the ratcheting 
of actin, and the $\sigma$-term accounts for acto-myosin contraction. 
In Eq.~(\ref{peqsimple}), the terms $D_p\Delta\bp$ and $-\tau_1^{-1}\bp$ describe diffusion 
of actin and its degradation (depolymerization) in the bulk of the cell, respectively.
The term $-\beta \nabla \rho$ describes the creation of actin polarization at the cell membrane 
(directed normal to the interface) with polymerization rate $\beta$.
The contribution $-\tau_2^{-1}(1-\rho^2)\bp$ assures a vanishing 
polarization outside of the cell (where $\rho=0$). Finally, 
$-\gamma\left[(\nabla\rho)\cdot\bp\right]\bp $ models the front-rear
symmetry breaking induced by motors. For details we refer to \cite{ziebert2011model,ebook}. 

Since motile cells are rather thin (typical lamellipodium thicknesses are $200$ nm)
the model is effectively two-dimensional, i.e.~height averaged.
In addition,  keratocyte cells are known to preserve their contact area 
with the substrate. To describe this conservation of the cell's contact area,
%-- to which we refer to as the 2D volume in the following --
we introduced the following global constraint 
\begin{equation}\label{delta1}
\delta=\delta_V=\frac{1}{2}+\mu_V\left[V(t) - V_0\right].
\end{equation} 
Here $\mu_V$ is the stiffness of the constraint and the term in brackets 
is the difference between the current area (or 2D volume)
$V(t)=\int\hspace{-1mm}\rho(t)\,dx\,dy$ and the prescribed area $V_0$.
Note that, to avoid confusion, in the following {\it area} always corresponds to the 2D area
of the cell's cross-section (corresponding in a 3D description to the cell's volume), 
while the membrane refers to the surface, i.e.~circumference, of this cross-section 
(corresponding in a 3D description to the cell's surface area). 

The position of the interface -- which is identified with the cell membrane -- 
can be defined in the model by the contour at $\rho=\frac{1}{2}$.
However, %in current models %[cf.~Eq.~(\ref{simple}) and (\ref{reqsimple0})] 
this interface is not an appropriate description for a cell membrane:
it has neither membrane tension nor bending energy, 
but rather an (artificial) wall energy ($\propto\sqrt{D_\rho}$) that is
related to the Ginzburg-Landau-type free energy of the phase field, cf.~Eq.~(\ref{simple}).

Even more important in the context of cell motility is the fact that
membrane tension counteracts the polymerization force of the actin filaments:
polymerization rate and hence the cell's velocity decrease
as a function of the counteracting force, as established theoretically 
on a single filament level by the Brownian ratchet model \cite{Oster:ratchet,MogilOster1}.
Although studies of single/few actin filaments polymerizing against a load are very 
difficult, this effect could also been established experimentally \cite{Kovar,Dogteromactin,Bibette}.
The membrane tension feedback on actin polymerization possibly not only leads
to a change in the overall velocity of the cell, but also 
to a global feedback on the actin organization and a change in the overall shape of the cell.

\section{Membrane tension as a counteracting force to polymerization}
\label{secmt}

We will first focus on the effect of membrane tension on actin polymerization
within the whole cell model described in the last section.
To this effect, we remove the -- artificial -- wall energy of the phase field,
and add the restoring force of the membrane counteracting polymerization.
%and add the resistence due to tension to the actin polymerization.
For simplicity, we keep the simple volume conservation and ignore at first the effect 
of tension on the phase field, a limit corresponding to a strongly adhering cell
that  keeps its contact area constant.
The effect of tension on the phase field is added and studied in the next section.

%{\bf Removal of wall energy.} 

The wall energy of the phase field potential can be removed -- to leading order in the interface width -- via
addition of the following term to the phase field equation \cite{Folch1,biben}: 
$\p_t\rho=\ldots+D_\rho c |\nabla\rho|$, where $c=c(x,y)$ is the curvature of the interface.
The latter can be calculated from the local normal unit vector, which is determined by the phase field like 
$\bn(x,y)=\frac{\nabla\rho}{|\nabla\rho|}$, via the geometric identity $c=-\nabla\cdot\bn$. 
Note that with the given definition, the normal vector points to the inside of the cell.

%{\bf Membrane tension and feedback on polymerization.}
Second, we introduce the membrane tension $\zeta(x,y)$. 
In the Helfrich picture, the membrane energy reads \cite{biben}
\beq\label{bend2d}
E_{mb}=\int\zeta|\nabla\rho|\,dx\,dy+\frac{b}{2}\int c^2|\nabla\rho|\,dx\,dy\,.
\eeq
The first term implements the surface area constraint 
(circumference in 2D), where $\zeta$ is the membrane tension, 
i.e.~the Lagrangian multiplier associated with the constraint.
The second contribution is the bending energy with the curvature $c$ already introduced above and $b$ 
the corresponding bending modulus. 
The restoring force resulting from this energy has been calculated in \cite{Cantat_bookchap} and reads
\beq\label{restoreforce}
\mathbf{F}_{mb}%(\zeta,c)
=\left[
\zeta c{\bn}
-b\left\{\frac{c^3}{2}+{\bt}\cdot\nabla\left({\bt}\cdot\nabla c\right)\right\}{\bn}
+({\bt}\cdot\nabla\zeta){\bt}
\right]|\nabla\rho|\,.
\eeq
Herein, the first term is the effect of the tension, the second one the contribution 
from bending and the last one arises from possible variations in tension along the membrane 
(similar to the Marangoni effect in thermal convection \cite{chandrasekhar2013hydrodynamic}).
 
Since the polymerization is normal to the membrane, we use only the normal contribution $F=\bn\cdot\mathbf{F}_{mb}$.
In addition, one can estimate that the contribution to the restoring force from bending ($\propto b$) is 
negligible, see \ref{appPar}. Hence we simply obtain that at the membrane 
(where $|\nabla\rho|$ is non-vanishing)
the  restoring force in normal direction is given by $F =\zeta c$: 
the force counteracting polymerization 
is proportional to the tension and the local curvature.

The simplest way to determine the value of the tension $\zeta$
is to assume that it is related to the overall relative excess circumference  
of the membrane (in 3D: surface area) 
via\footnote{Since $\nabla \rho$ is nonzero only at the interface, 
$\int\hspace{-1mm}|\nabla\rho|(t)\,dx\,dy$ is (proportional to) the cell's circumference. }
\beq\label{zeta}
\zeta=T\delta A= T\frac{A(t)-A_0}{A_0}\,,\,\,{\rm where}\,\,A(t)=\int\hspace{-1mm}|\nabla\rho|(t)\,dx\,dy\,.
\eeq 
Here $T$ is the membrane's compressibility modulus 
and $A_0$ is %=\int\hspace{-1mm}|\nabla\rho_0|(t)dx dy$ 
the cell's circumference in a reference state. For the latter we chose the stationary, round, non-moving state,
as it has the smallest circumference. 

%We can mentioned that we also tried dynamical versions.

Finally, we have to account for the feedback the membrane tension provides on the actin
polymerization dynamics.
In the simple model, the term $-\beta\nabla\rho$ in the equation for $\bp$ described
that at the position of the membrane (where $\nabla\rho$ is nonzero), actin is created
with constant rate $\beta$ in the normal direction\footnote{Note that $\nabla\rho$ is negative
when measured  from inside the cell, hence the `-'sign.}.
Here, we make use of the ratchet-like process of actin polymerizing against the membrane,
where it had been shown that the polymerization rate $\beta$ decreases exponentially
(in the simplest case) with the force
\beq
\beta(F)=\beta\exp\left(-\frac{aF}{k_BT}\right)=\beta e^{-f_0\zeta c}\,,
\eeq
where $k_BT$ is the thermal energy and $a$ is the size of the actin monomer
\cite{Oster:ratchet,MogilOster1}, which can be absorbed in the constant $f_0$.
Note that the parameter $\beta$ associated with
actin polymerization is not the free polymerization rate\footnote{
Since $\beta$ is not the free polymerization rate, for negative curvature (locally concave shape)
there could be a slight acceleration of actin polymerization until the free polymerization
rate is reached. We neglect this effect here, since it is small and stationary cell shapes
are almost exclusively convex, by considering only $c>0$ in the exponential.
In the wall energy correction term, %$D_\rho c|\nabla\rho|$, 
however, both curvature signs have to be considered.
}, 
but the rate reduced by the offset tension present in the reference state, cf.~Eq.~(\ref{zeta}). 
Correspondingly,  the propulsion strength, $\alpha\bp$, decreases
too upon an increase in membrane tension, which is 
due to the smaller amount of overall actin polarization $\bp$.

The generalized model  then reads
\begin{eqnarray} 
\label{reqwm1}
\partial _t\rho\hspace{-2mm}&=&\hspace{-2mm}
D_\rho\Delta\rho-\rho(1-\rho)(\delta-\sigma|\bp|^2-\rho)
+D_\rho c |\nabla\rho|
%-\frac{\delta F_b}{\delta\rho}
-\alpha\bp \cdot\nabla\rho\,,\quad\\
\label{peqwm1}
\partial_t\bp\hspace{-2mm}&=&\hspace{-2mm}D_p\Delta\bp
-\beta e^{-f_0\zeta c}\,\nabla\rho\,
%-\beta\exp(-f_0\zeta c)g(\nabla\rho)\,
-\tau_1^{-1}\bp   
-\tau_2^{-1}(1-\rho^2)\bp 
-\gamma\left[(\nabla\rho)\cdot\bp\right]\bp\,,\nonumber\\
\end{eqnarray}
with $\delta$ given by Eq.~(\ref{delta1}) 
and $c=-\nabla\cdot\bn=-\nabla\cdot\left(\frac{\nabla\rho}{|\nabla\rho|}\right)$ 
the curvature as described above.
For numerical reasons, the normal vector and the curvature can only be calculated in a `tube' around the 
interface (using a threshold value for $\nabla\rho$). Upon multiplication with $|\nabla\rho|$ the
respective terms nevertheless lead to smooth contributions at the interface, where they are needed. 
%The normal vector entering the curvature is calculated for the region where $|\nabla\rho|$
%exceeds a threshold value (typically we used $0.01$), corresponding to a tube around the interface. 
%Outside the effect is negligible due to the presence of $\nabla\rho$ in both terms involving $c$. 
%{\color{red} Benjamin, do you still have the smoothened transition or just $c=0$ outside?}
%({\it We should add where we calculate c, i.e. for $|\nabla\rho|>n_c$ for a threshold value $n_c$}).

{\bf Results for steady moving cells.} 
%We here focus on the feedback 
%of the membrane's restoring force on the actin polymerization.
%Volume is conserved, the wall energy switched off via the correction term, and 
%bending in the phase field is not present.
We investigated the behavior of the following  quantities that can be easily measured in experiments: 
the aspect ratio of the cell as a measure for the shape change, 
the cell's velocity and
its relative excess circumference. 
Note that the latter is not directly restricted via the phase field --
only the volume is conserved -- but only indirectly via the feedback provided on the polymerization rate. 
The aspect ratio has been characterized as previously \cite{ziebert2011model}, 
by determining the ratio $h$ of the eigenvalues of the variance matrix
$I_{ij}=\int(x_i-x_i^c)(x_j-x_j^c)\rho\,dx\,dy$, 
where $\mathbf{r}^c=\int {\bf r} \rho\,dx\,dy$ is the center of mass of the cell.
Since a round cell has $h=1$, we treat $h-1$ as a measure for the deviation from a circle.

\begin{figure}[t!]
        \centering
                \includegraphics[width=0.9\textwidth]{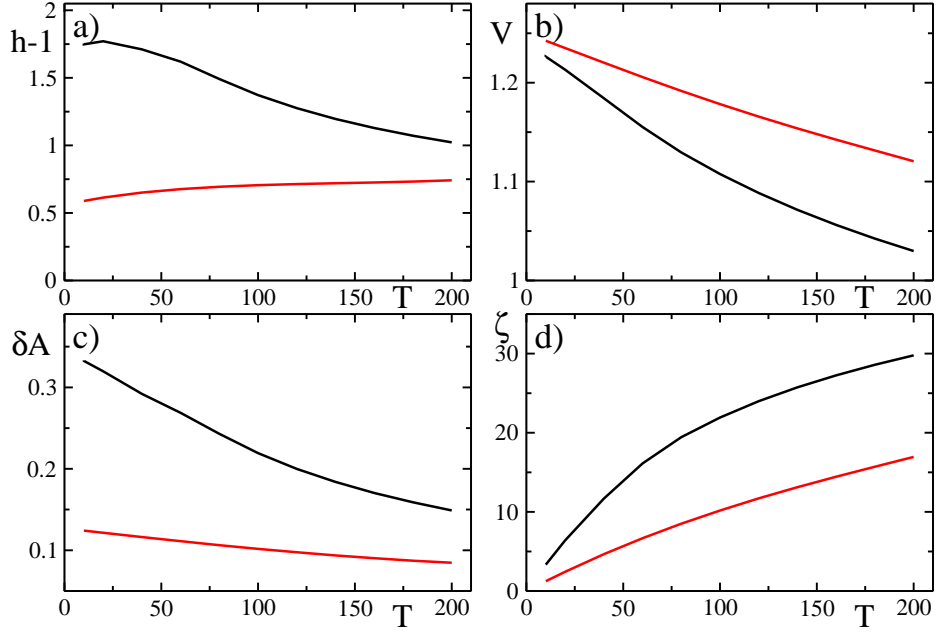}
\caption{
The aspect ratio's deviation from round shape, $h-1$ (a), 
the center of mass velocity $V_{\rm com}$ (b), 
the relative excess circumference   $\delta A$ (c),
and the membrane tension $\zeta=T\delta A$ (d), 
for a keratocyte-like cell 
(black curves; $\alpha=2$, $\beta=2$, $\sigma=1.2$, $\gamma=0$), 
and for a fibroblast-like cell 
(red curves; $\alpha=3$, $\beta=1.5$, $\sigma=0.9$, $\gamma=0.3$) vs membrane compressibility constant $T$.
Remark: the volume is conserved by better then 1\% in both cases.
%{\color{red} I will add part d) for the tension 
%which is $\zeta=T\delta A$ and hence increasing (from $\simeq$7.5 to 30 for the black curve), 
}
        \label{fig1}
\end{figure}

Figure~\ref{fig1} displays a) the aspect ratio's deviation from the round shape, $h-1$, 
b) the center of mass velocity $V_{\rm com}$,
c) the relative excess circumference   $\delta A$
and d) the membrane tension $\zeta$, %=T\delta A$,
as a function of the membrane's compressibility modulus $T$. 
Shown are results for two different cells, one of keratocyte shape (black curves)
and one of a more fan-like, fibroblast shape (red curves), cf. also figure 2. 
In case of the first cell type, the aspect ratio expectedly decreases by increasing $T$:
at the sides the curvature is highest, and hence the restoring force leads to a more rounded shape.
In contrast, for the second cell type, the aspect ratio increases. This is due to the
fan-like, triangular shape of the cell, where membrane tension not only reduces the extension
normal to the direction of motion, but also in direction of motion.  
The velocity is not substantially affected (in the $10-20\%$ range),
in accordance with experiments \cite{Fuhs_Kaes,Keren_mb}. %({\color{red}verify latter ref}).

The relative excess surface decreases in both cases, cf.~Fig.~\ref{fig1}c), which shows that in the
limit of high $T$, the surface (circumference in 2D) should become rather well conserved,
even without explicit inclusion into the phase field equation. 
Although the relative excess surface decreases, the overall tension $\zeta=T\delta A$ increases,
cf.~Fig.~\ref{fig1}d),
as it should. Experimentally, also the membrane tension is accessible by pulling membrane tethers from 
moving cells, as has been recently studied in \cite{Keren_mb}.

%{\it remark: We could also show cuts through the cell in the direction of motion and normal to it.
%There one can see that the membrane restoring force is very high (of order 4) on the sides,
%below 0.1 at the front, which is rather straight, and a bit higher (0.2) at the slightly curved back. }

\begin{figure}[t!]
        \centering
                \includegraphics[width=0.6\textwidth]{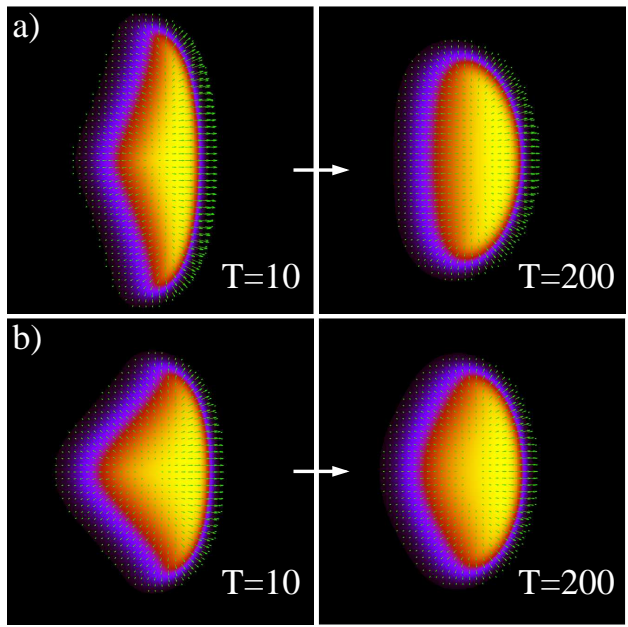}
\caption{The shapes for a) a keratocyte-like cell 
($\alpha=2$, $\beta=2$, $\sigma=1.2$, $\gamma=0$) and
b) a fibroblast-like cell 
($\alpha=3$, $\beta=1.5$, $\sigma=0.9$, $\gamma=0.3$)
for small ($T=10$)  and large ($T=200$)  membrane's compressibility modulus, cf.~also
the curves in Fig.~\ref{fig1}.}
        \label{fig2}
\end{figure}

Figure~\ref{fig2} shows the respective shape changes, comparing low and high membrane extensibility moduli
(end hence tension values) for the two kinds of cells.

\begin{figure}[t!]
        \centering
                \includegraphics[width=\textwidth]{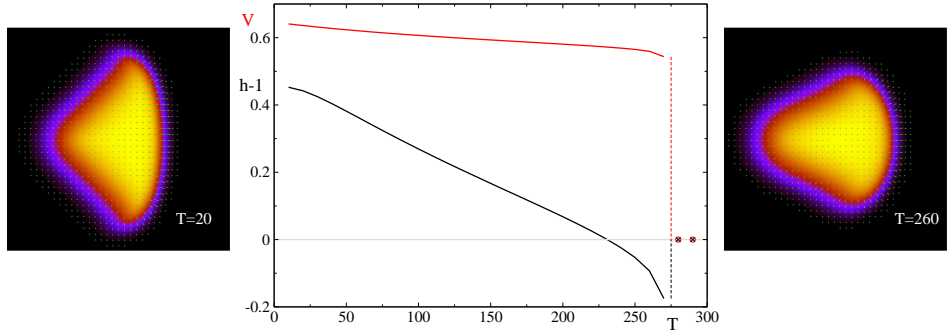}
\caption{Evolution of the aspect ratio's deviation from round shape, $h-1$, 
and center of mass velocity $V_{\rm com}$ 
for mildly driven cells (parameters: $\alpha=1.3$, $\beta=1.3$, $\sigma=1.4$, $\gamma=0$).
The tension modulus $T$ was increased step by step and the velocity and aspect ratio measured
after the cell had equilibrated to its steady state. 
At a certain critical value $T_c$ the cell stops.
The snapshots at the left and right show the cell's shapes for 
small (negligible) tension and for the highest tension value
where the cell still is able to move.
 }
        \label{fig3}
\end{figure}

{\bf Arrest of cell motion by  increased  tension.}
Next we have studied whether the onset/cessation of motion is affected by membrane tension.
Mildly driven cells (i.e.~cells with not too large values of the propulsion parameter $\alpha$, 
as well as $\sigma$ and $\gamma$) can indeed be stopped by increasing the compresibility modulus $T$
and hence the tension.
This effect has been seen in several experiments, e.g.~in \cite{Houk} for leukocytes,
and is captured by our model as shown in Fig.~\ref{fig3}.
Interestingly, while without membrane tension both velocity 
and aspect ratio exhibit a jump at the arrest of motion, cf.~\cite{ziebert2011model},
with tension present the aspect ratio decreases continuously and
the cell becomes even slightly stretched in the 
direction of motion, before finally being stopped for even higher tension. 
The reason is that the propulsion is predominantly due to the front-rear asymmetry. 
Increasing $T$ strongly affects the sides (i.e.~the aspect ratio  decreases),
but the feedback on polymerization is small at the front (and typically also at the rear) 
since curvature is small there. Consequently, the jump in the velocity is decoupled from 
the jump in the aspect ratio, due to tension\footnote{
Note that the study in Ref.~\cite{ziebert2011model} indicated that  
there is not a simple relation between the aspect ratio and the center of mass 
velocity, e.g.~when varying the cell's contact area (volume in 3D); 
here the same is true even for fixed contact area, but for different values of tension. 
Hence again, the aspect ratio is not necessarily related in a simple way to the cell's velocity.}. 

Experiments indicated a direct relation between membrane tension and the actin 
pushing force: reducing the latter results in a decrease in membrane tension,
as found by treating keratocytes with blebbistatin and jasplakinolidine, 
leading to a rapid cessation of the actin assembly \cite{Keren_mb}.
Ref.~\cite{Houk} found the same, namely that membrane tension increases upon stronger 
leading-edge protrusion.
This relation can be directly inferred from  our model: namely, it is the propulsion by actin $\propto\alpha$ 
(and caused by actin polymerization $\propto\beta$) that induces an increase 
in the excess area $\delta A$ (even in the non-motile case) and hence
in tension $\zeta=T\delta A/A_0$, cf.~Eq.~(\ref{zeta}).
 
Another interesting experiment performed in Ref.~\cite{Keren_mb} was the 
fusion of a moving keratocyte with a giant unilammelar vesicle (GUV)
to increase the available membrane area. 
The cross-sections of the cells were found to become larger 
(i.e.~volume entered from the third dimension) with higher aspect ratios,
while the velocity and tension remained the same. This behavior too is
in accordance with our  model: after fusing a cell with a GUV,
both the contact area $V_0$ and the circumference $A_0$ will increase due to spreading. 
We already studied the behavior of velocity and aspect ratio as a function of 
the cell's contact area in \cite{ziebert2011model},
and found that (keeping all other parameters fixed)  
for not too large cells the aspect ratio increases with area
while the velocity remained practically unchanged.
Membrane tension will not change this scenario due to the concomitant increase in $A_0$
(the restoring force may decrease a little, since it is proportional to curvature,
but the leading edge is rather straight anyways).

In Ref.~\cite{Keren_mb} it was also observed that  
an increase in cellular adhesion to the underlying substrate increased the membrane tension. 
Again, this is consistent with our model since the propulsion force is proportional 
to the number of adhesive ligands $A$, $\alpha\simeq\alpha_0 A$, see 
Ref.~\cite{ziebert2013effects} for a generalization of the model including explicit adhesion dynamics.  
Consequently, upon increased adhesion the cell can spread more efficiently, 
thereby increasing the relative excess area and consequently the tension $~\delta A T$.

Finally, Ref.~\cite{Keren_mb} found that a decrease of myosin contraction leads to higher tension.  
This is the only trend not (yet) captured in our model: if the parameters associated with the activity of 
motors ($\sigma$ or $\gamma$) are decreased, we do obtain more round
shapes, but this is not due to an increase in tension.
The reason for this discrepancy might well be that the implementation of the action of motors   
is still too oversimplified: we neither implemented explicit motors, nor tensorial active stresses, 
nor the contractile bundle at the rear present in keratocytes.

\section{Contact area vs.~contour conservation}
\label{areavscontour} 

There arises the question, which quantity should be conserved in an effective
2D model of a cell. For a 3D cell this is rather clear: 
the volume is conserved since the cytoplasm is incompressible, 
and the membrane area too since the membrane's compressibility modulus is very high,
implying almost perfect inextensibility (note, however that there are cells with membrane folds to buffer
surface area, see also below). 
In contrast, for a height-averaged 2D model as ours, both the contact area could vary
-- e.g.~the cell retracts to the third dimension by reducing its spreading -- 
and the circumference.

One can describe  this effect on a phenomenological level 
by considering the phase field parameter $\delta$ to depend on both the contact 
area and the length of the circumference:
\beq\label{deltafull}
\delta=\frac{1}{2}+\mu_V\left[V(t) - V_0\right]
+\mu_AT\left[A(t)-A_0\right].
\eeq
Here, we tuned $\mu_V$ and $\mu_A$ in such a way that for a moderate 
value of the compressibility modulus, $T=50$, both contributions 
are of same order (for a cell of specific size; we typically used cell's of radius $r_0=15$).
Hence $T=0$ corresponds to pure contact area conservation, while
the limit $T\gg50$ leads to a dominating conservation of the circumference.

\begin{figure}[t!]
        \centering
                \includegraphics[width=0.8\textwidth]{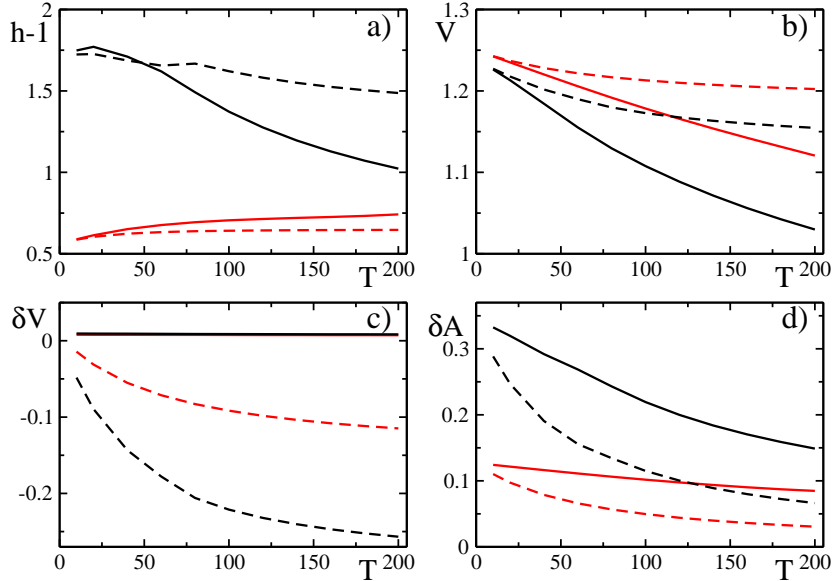}
\caption{Shown are the aspect ratio's deviation from round shape $h-1$ (a), 
the center of mass velocity $V_{\rm com}$ (b), 
the excess contact area $\delta V$ (c),   
and the relative excess circumference $\delta A$ (d) for two cell types 
as a function of the membrane's compressibility modulus $T$:  
for a keratocyte-like cell 
($\alpha=2$, $\beta=2$, $\sigma=1.2$, $\gamma=0$) with area conservation (black solid) 
and with combined area-circumference conservation (black dashed),
as well as for a fibroblast-like cell 
($\alpha=3$, $\beta=1.5$, $\sigma=0.9$, $\gamma=0.3$), again with area conservation (red solid) 
and with combined area-circumference conservation (red dashed).
Note that the solid curves are identical to those in Fig.~\ref{fig1}.}
        \label{fig4}
\end{figure}

We investigated  the model described  by Eqs.~(\ref{reqwm1}) and (\ref{peqwm1}),
with  the parameter $\delta$ replaced by Eq.~(\ref{deltafull}). 
Figure~\ref{fig4} displays  the aspect ratio, the center of mass velocity, the contact area $\delta V$, 
and the circumference $\delta A$, as a function of the membrane's compressibility modulus $T$.
Expectantly, with combined area-circumference conservation, 
the circumference is better conserved at the cost of the contact area 
conservation [see panels c) and d)]. 
Figures~\ref{fig4}a) and b) show that the effect of tension on both the aspect ratio and the velocity 
are much smaller than when only  area conservation constraint is imposed. 
Hence interestingly, although one would naively expect a decrease 
of speed from the exponential decrease of polymerization with tension,
the speed remains almost unchanged due to the global shape dynamics:
one can not conclude the overall speed just from the local polymerization rate.
In case of only contact area  conservation, the  velocity decreased by 10-20\%, 
cf.~also Fig.~\ref{fig1}. In contrast, if the  circumference conservation is important,  
the overall decrease is below 5\% and probably not even measurable in experiments.
Note that keratocytes are known to have only transient, weak adhesion and high membrane 
tension \cite{Keren_mb}, hence circumference conservation should be important. 
Indeed both experiments \cite{Fuhs_Kaes,Keren_mb} did not see an effect 
of membrane tension on the cell's velocity. We are not aware of such experiments
on fibroblasts, but given their higher adhesion and lower tension, the study in the previous
section suggests that their velocity might be affected in the 10-20\% range.

Thus, our  study indicates  that the exchange of contact  area and  membrane circumference with
the third dimension is crucial for  understanding  the influence of membrane tension on the shape
and speed of cells. Obviously, this problem can only be properly addressed
in truly three-dimensional models which are underway \cite{CatesNCom}.
Note that the membrane's curvature in this third direction is typically much higher 
than the in-plane one (due to the thinness of the lamellipodium) and might be the dominating 
curvature \cite{Verkhovskycontactangle}.
Another complication arises due to the existence of plasma membrane reservoirs.
While these reservoirs  seem not to be present in keratocytes \cite{Keren_mb}, they are known to exist
for neutrophils (where they are important for phagocytosis \cite{HerantDembo06}, i.e.~the 
uptake of micron sized objects),
as well as in fibroblasts. The percentage of area buffered/stored can be between
1\% for strongly adhering, up to 10\% for spreading fibroblasts \cite{Gauthier_Sheetz,Raucher_buffer}.

\section{Effect of bending rigidity \& perturbing moving cells} %: obstacles \& tunnels}

The bending rigidity contribution to the force opposing actin polymerization
is typically negligible compared to the contribution due  to tension (cf.~Section \ref {secmt} and Appendix).
Nevertheless, bending may affect the shape of the cell directly when its
contribution to the phase field equation becomes relevant.    
From the typical 2D radii of curvature (of order $20\,\mu{\rm m}$) and typical membrane rigidities, 
one expects only moderate effects for stationary moving cells
-- mostly at the wings where the curvature is the highest.
However, bending may well be relevant for the dynamics of cells and for the response
to external perturbations, e.g.~via modulations of the substrate properties as studied
in \cite{ziebert2013effects,lober2014modeling} or by obstacles.

%overall shape, especially and also lead to better stability/cohesion of the cell's interface when externally
%perturbed.
For vesicles,
the bending energy can be introduced similar as the phase field energy in Eq.~(\ref{Fdef}), 
by considering the already introduced function
$\lambda=D_\rho\Delta\rho-f^\prime (\rho)$.
Helfrich-Willmore theory \cite{HelfWilm} then implies the bending energy to be
\cite{VoigtJRSI}
\beq\label{Fbphase1}
E_b=\frac{6\sqrt{2}}{D_\rho\sqrt{D_\rho}}\,b\int \lambda^2\,dx\,dy\,.
\eeq
In the sharp interface limit, see Ref.~\cite{HelfWilmConv}, this expression
tends to $b\int C_M^2 dx\,dy$, where $C_M$ is the mean curvature
and $b$ is the bending modulus.
%(similarly, $F_S$ then tends to a surface energy $\propto\sqrt{D_\rho}$).
%The additional term $-\frac{\delta F_B}{\delta\rho}$ in the dynamic phase field equation 
%can be evaluated as
%\beq
%\frac{\delta F_B}{\delta\rho}=b^*\Delta\lambda-\frac{b^*}{D_\rho}f^{\prime \prime} (\rho)\lambda\,
%\eeq
%where $b^*=\frac{2b}{\sqrt{D_\rho}}$.
However, this approach can only be used in the `advected field' case,
i.e.~if the shape of the phase field $\rho$ across the interface 
stays (close to) a $\tanh$-profile  all the time. This is the case
for vesicles in Stokes flow, and also in the active gel hydrodynamics approach
recently studied \cite{VoigtJRSI}, since there are no forces normal to the interface.
In the model for a moving cell considered here, however, the active terms (proportional to $\alpha$ and $\sigma$)
{\it are} normal forces and  deform  the interface away from the radial $\tanh$-shape.
We checked  the overall bending energy using Eq.~(\ref{Fbphase1}) for stationary round cells. 
We obtained that for $\alpha\gtrsim0.5$ -- already well below values needed to induce motility --
the bending energy not only increased in value, but,  in addition,  did not  display anymore
the correct behavior $E_b\propto\frac{b}{R}$. Hence in Eq.~(\ref{Fbphase1}) 
one would have to correct for the terms perturbing the radial profile of the phase field, 
which is neither a straightforward task nor very intuitive.  

We therefore applied a purely 2D formulation for the bending energy \cite{Cantat_bookchap} 
given by Eq.~(\ref{bend2d}), i.e.~used
\beq
E_{b}=\frac{b}{2}\int c^2|\nabla\rho|\,dx\,dy\,,
\eeq  
leading to the following contribution in the phase field\footnote{
Since the bending contributions scale with the radius of curvature like $1/R^3$ and rapidly grow towards
the inside of the cell, for numerical reasons one has to decrease $D_\rho$ slightly 
(increasing the sharpness of the interface) and also the tube width around the interface
where the terms are calculated.}
\beq\label{restoreforcepf}
\partial_t\rho= \ldots + b\left\{\frac{c^3}{2}+{\bt}\cdot\nabla\left({\bt}\cdot\nabla c\right)\right\}|\nabla\rho|\,.
\eeq

\begin{figure}[t!]
        \centering
                \includegraphics[width=0.99\textwidth]{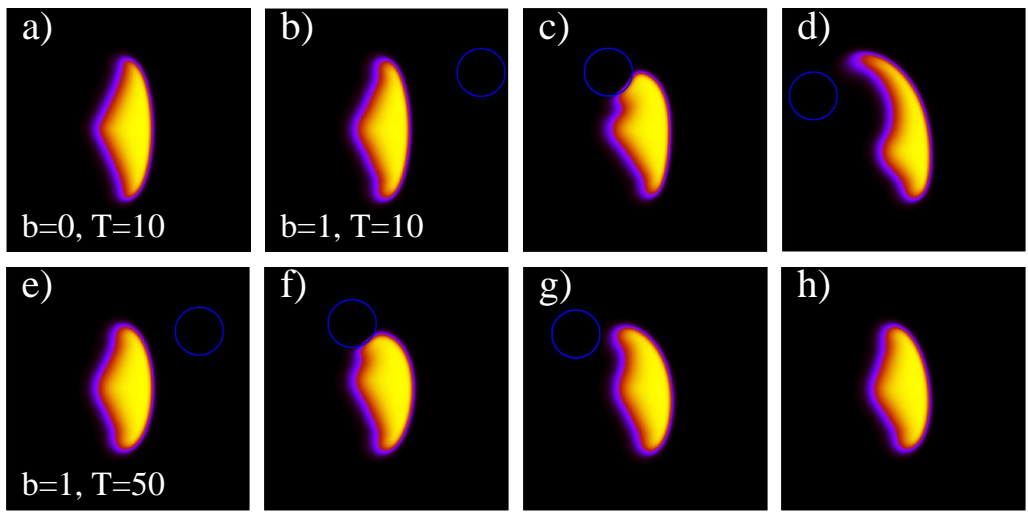}
\caption{Sequences of snapshots illustrating the collision between a moving cell 
and an obstacle. Panel  a) shows a steady moving cell with low tension ($T=10$) and no bending. 
Panels b)-d) display the collision of a cell with the same tension and bending rigidity $b=1$ with a round obstacle. 
From the shape of the steady moving cell b) in comparison to a) 
one can infer that bending leads to a slight rounding of the wings and the back. 
For such a `floppy' cell of low tension, the collision leads to a strong deformation of the
wing hitting the obstacle.
Parts e)-h) show a collision of a cell with higher tension $T=50$ and same bending rigidity,
with an obstacle. The cell is much less deformed, recovers its initial shape more rapidly
and is slightly deflected from its initial direction of motion.
The collision events have been traced in the moving frame of the cell.
Other parameters: $\alpha=\beta=1.3$, $\sigma=0.6$, $\gamma=0.4$; $D_\rho=0.5$ in place of $D_\rho=1$
which was used in the other figures.
}
        \label{fig5}
\end{figure}

%Since the bending energy contribution is typically  small and can be easily overshadowed by other effects, 
%we used Eqs.~(\ref{reqwm1}) and (\ref{peqwm1}) with the contact area  
%conservation constraint only and for $f_0=0$, i.e.~neglecting again the membrane's feedback on
%polymerization. 
As expected, for steady moving cells the effect of bending is not very noticeable, 
except for very low tension values and rather strongly elongated cells.
Figures ~\ref{fig5}a) and b) show steady moving cells with $T=10$, and $b=0$ and $1$, respectively.
One can see that bending leads to a rounding of the wings and slightly also of the back. 
%This behavior persists  for rigidities up to $b=5$, which is already a high value accounting also for the stabilizing
%effect of an actin cortex underneath the membrane. 
%Only the wings are slightly rounder for rather strongly elongated cells
%(e.g.~$\sigma=0.6$, $\gamma=0.4$), which however is not clearly visible in the aspect ratio. 

%To study the effect of both bending and membrane tension on reshaping cells,
We also investigated collisions of cells with rigid round obstacles, to test whether 
perturbations externally imposing a curvature lead to stronger effects.
For this purpose, the obstacle was implemented by a second stationary phase field $\rho_o(x,y)$
(that had been relaxed towards a radial $\tanh$-profile of equal width as the stationary cell)
and coupled to the cell via a steric interaction energy (as developed previously for multiple cells
\cite{LoeberSciRep})
\beq
E_{\rm steric}=\frac{1}{2}k\rho^2\rho_o^2\quad\rightarrow\quad\partial_t\rho= \ldots -k\rho_o^2\rho\,.
\eeq 
For the interaction strength we chose $k=10$, typically, to prevent overlap.
Figures ~\ref{fig5}b)-d)  shows a collision with an obstacle of radius $R_o=10 \mu m$
for a cell with low tension, $T=10$, and high bending rigidity, $b=1$.
For comparison, Figures ~\ref{fig5}e)-h) display a cell with the same bending rigidity but higher tension $T=50$.
One can clearly see that the cell with low tension becomes  very deformed in spite of its bending rigidity,
while the second one much less. The second cell relaxes rapidly to its initial shape,
albeit deflected in its direction of motion, cf.~also Suppl.~Movies 1 \& 2.  

Finally, we also studied cells encoutering a narrow channel, a situation
increasingly studied experimentally \cite{Spatzchannel}, cf.~also \cite{rhoda}.
Expectantly, for large enough channel widths the cells pass, while narrow channels stop them 
at the entrance.
Suppl.~Movies 3 \& 4 focus on the intermediate behavior: Movie 3 shows a cell with low tension, $T=10$,
that is able to go through the channel albeit being slowed down.
Movie 4, in contrast, shows a cell with high tension, $T=50$, encountering the same channel.
It is less deformable and hence does not pass.

\section{Conclusions}

We investigated the consequences of a variety of physical mechanisms associated with the cell membrane, 
such as the feedback of membrane tension on actin polymerization, 
contact area vs.~circumference conservation, and membrane bending stiffness,  
on the shape and motility of moving cells.
These questions are important in the context of cell motility, since especially the membrane tension acts as a global
mechanical feedback that may constitute a conduit for rapid -- note that tension relaxes on
the time scale of few milliseconds -- information transfer across the cell.
In addition, physical perturbations of cells,  by e.g.~collisions or substrate modulations, 
are increasingly investigated. The overall conclusion is that  the membrane plays an important role in preserving the
integrity of the cell as well as for its dynamical response.   

We have found that the dominant effect of the membrane on the cell's motility 
is its feedback via tension on the ratcheting of actin. 
This is in accordance with the finding by Lieber {\it et al.} \cite{Keren_mb} 
that for keratocytes, membrane tension is dominated by the cytoskeletal forces. 
Since keratocytes have rather high tension compared to other moving cells, the most appropriate
model for this cell type is the one with combined contact area and circumference conservation,
developed in section \ref{areavscontour}.
Accordingly, bending only leads to small corrections, mostly during cellular reshaping events. 

Interesting future aspects of membrane effects include generalizations
to 3D models of cells. There, the high out-of-plane curvature at the tip of the
lamellipodium may dominate the behavior \cite{Verkhovskycontactangle}. 
Another important aspect is related to the fact that the tension originates 
not only from the lipid  bilayer membrane, but also from the actin cortex directly underneath, 
adhering to the membrane.
A loss of the membrane-cytoskeleton adhesion then can lead to blebbing \cite{Sheetzbleb},
constituting another means of motility, which has not yet been modeled in a whole cell, dynamical model.

%The following can be still done if time permits,...\\
% Effect of membrane  on stick-slip dynamics? Would need to reintroduce
%two more equations, but could be interesting.\\



\section*{Acknowledgments}
B.W. and F.Z. acknowledge funding from the German Science Foundation (DFG) via project ZI 1232/2-1. I.S.A. was supported by the US Department of Energy (DOE), Office of Science, Basic Energy Sciences (BES), Materials Science and Engineering Division. The numerical work was in part performed on the Northern Illinois University GPU cluster GAEA.

\appendix
\section{Parameters and estimate of tension vs.~bending}
\label{appPar}

The typical scales in our model are chosen to be seconds for time, microns for length
and piconewtons for force (see \cite{ziebert2011model} for details). 

The force given in Eq.~(\ref{restoreforce}) was derived for vesicles 
\cite{Cantat_bookchap} and is a volume force (entering the Stokes equation),
i.e.~it has units of ${\rm N/m^3}$. To obtain  the force acting  on an actin filament, 
we hence have to multiply by the volume of an actin monomer, roughly $a^3=(10\,{\rm nm})^3$.
In the feedback on the polymerization rate, in the expression $\exp\left(\frac{aF}{k_BT}\right)$ one also has  
 $a=10\,$nm and $k_BT\simeq 4\,{\rm pN nm}$.

The membrane's compressibility modulus is of order $T=0.1{\rm J/m^2}$ \cite{Fletcherrev},
leading to $10^5\,{\rm pN}/\mu{\rm m}$. For numerical reasons we varied $T=0-500$. 
The tension $\zeta$ should be of the same  order \cite{Fletcherrev}; 
it has been recently measured in \cite{Keren_mb} by pulling membrane tethers
from moving keratocytes, and was found  to be of the  order of $250\,{\rm pN}/\mu{\rm m}$.
The force due to membrane tension 
can be hence estimated to be $\zeta c |\nabla\rho|a^3\simeq 0.01-1$\, pN, depending on whether one
uses the typical phase field interface width ($\simeq 1\,\mu{\rm m}$) or rather 
a real membrane thickness (5\,nm).
%\beq
%\zeta c |\nabla\rho|a^3=1\frac{N}{m}\times\frac{1}{1\mu m}\times\frac{1}{10 nm}\times(10 nm)^3
%= 100 pN
%\eeq
%This would be a 'realistic' value, where the membrane is 10 nm thick, cf. 
%I used $|\nabla\rho|=1/(10nm)$. In the simulation,
%the phase field interface is spread much more (about length=1 i.e. 1 $\mu m$),
%which leads to 1pN then. 
Consequently, the prefactor $f_0$ in $\exp(-f_0\zeta c)$ occurring in Eq.~(\ref{peqwm1}) 
should be of order $0.01-1$, we typically used $1$.
 
In contrast, for the bending rigidity contribution in Eq.~(\ref{restoreforce}) we obtain 
$b c^3 |\nabla\rho|a^3=10^{-5}-10^{-7}{\rm pN}$, using a typical membrane bending modulus
of order of $b=100\,k_BT=0.4\,{\rm pN}\mu{\rm m}$.
In \cite{Keren_mb} also the bending rigidity has been directly measured for keratocytes 
and found to be slightly lower, $0.14\,{\rm pN}\mu{\rm m}$.
Thus, for the membrane's feedback on the polymerization rate,
bending is completely negligible vs.~tension.
In the phase field equation, however, the bending contribution enters directly 
%bending stiffness enters directly, 
i.e.~is of order $0.14-0.4$ in our units, we typically used $b=0-1$.

%\beq
%\kappa c^3 |\nabla\rho|a^3&=&100 k_BT\times\frac{1}{(1\mu m^3)}\times\frac{1}{10 nm}\times(10 nm)^3\nonumber\\
%&=&400pNnm\times\frac{1}{(1000 nm^3)}\times\frac{1}{10 nm}\times(10 nm)^3
%= 4\cdot10^{-5}pN
%\eeq
%which becomes even smaller when the phasefield interface is wider. 
%So in fact bending is negligible.

\bibliographystyle{elsarticle-num} 
\bibliography{ref}


%\begin{thebibliography}{00}


%\bibitem{}

%\end{thebibliography}

\end{document}